%% file: ms.tex
\documentclass[letterpaper, oldversion]{aa}  

\usepackage{natbib}\usepackage{graphicx}      
\bibpunct{(}{)}{;}{a}{}{,}

\newcommand{\kms}{\,km\,s$^{-1}$} \newcommand{\sqcm}{\,cm$^{-2}$}  
\newcommand{\lya}{Ly-$\alpha$} \newcommand{\qa}{\object{Q0841+129}}
\newcommand{\zabs}{$z_{\rm abs}$} 
\newcommand{\zdla}{2.47621} \newcommand{\zsub}{2.50620}
\newcommand{\zem}{$z_{\rm em}$}
\newcommand{\zqso}{2.49510$\pm$0.00003} 
\newcommand{\hi}{\ion{H}{i}}      \newcommand{\hw}{\ion{H}{ii}}    
\newcommand{\os}{\ion{O}{vi}}     \newcommand{\aro}{\ion{Ar}{i}}  
\newcommand{\cf}{\ion{C}{iv}}     \newcommand{\sif}{\ion{Si}{iv}}
\newcommand{\nf}{\ion{N}{v}}      \newcommand{\none}{\ion{N}{i}}
\newcommand{\cw}{\ion{C}{ii}}     \newcommand{\nw}{\ion{N}{ii}}
\newcommand{\sw}{\ion{S}{ii}}     \newcommand{\znw}{\ion{Zn}{ii}}  
\newcommand{\crw}{\ion{Cr}{ii}}   \newcommand{\sit}{\ion{Si}{iii}}  
\newcommand{\alt}{\ion{Al}{iii}}  \newcommand{\siw}{\ion{Si}{ii}}  
\newcommand{\fet}{\ion{Fe}{iii}}  \newcommand{\few}{\ion{Fe}{ii}}  
\newcommand{\oi}{\ion{O}{i}}      \newcommand{\ct}{\ion{C}{iii}} 
\newcommand{\mgw}{\ion{Mg}{ii}}   \newcommand{\hew}{\ion{He}{ii}}  
\newcommand{\lyb}{Ly-$\beta$}       

\begin{document}

\title{High-ion absorption in the proximate damped \lya\ system toward
  Q0841+129\thanks{Based   
  on observations taken under proposal IDs 65.O-0063(B), 70.B-0258(A),
  and 383.A-0376(A) with the Ultraviolet and Visual Echelle 
  Spectrograph (UVES) on the Very Large Telescope (VLT) Unit 2
  (Kueyen) operated by the European Southern Observatory (ESO) at
  Paranal, Chile.}} 
\author{Andrew J. Fox\inst{1,2}, C\'edric Ledoux\inst{1}, 
  Patrick Petitjean\inst{3}, Raghunathan Srianand\inst{4}, 
  \& Rodney Guimar\~aes\inst{5}}
\institute{
  European Southern Observatory, Alonso de C\'ordova 3107, Casilla
  19001, Vitacura, Santiago, Chile; afox@eso.org \and   
  Institute of Astronomy, University of Cambridge, Madingley Road,
  Cambridge, CB3 0HA, UK \and
  Institut d'Astrophysique de Paris, UMR7095 CNRS,
  UPMC, 98bis Blvd Arago, 75014 Paris, France \and
  Inter-University Centre for Astronomy and Astrophysics, Post Bag 4,
  Ganesh Khind, Pune 411 007, India \and
  Programa de Modelagem Computacional - SENAI - Cimatec, 41650-010
  Salvador, Bahia, Brazil}

\authorrunning{Fox et al.}
\titlerunning{High Ions in Q0841 PDLA} 
\date{Received 31 Mar 2011, Accepted 11 Jul 2011}

\abstract{We present VLT/UVES spectroscopy of the quasar Q0841+129,
  whose spectrum shows a proximate damped \lya\ (PDLA) absorber
  at $z$=\zdla\ and a proximate sub-DLA at $z$=\zsub, both lying
  close in redshift to the QSO itself at 
  $z_{\rm em}$=2.49510$\pm$0.00003. 
  This fortuitous arrangement, with the sub-DLA acting as a filter that
  hardens the QSO's ionizing radiation field,
  allows us to model the ionization level in the foreground PDLA, and 
  provides an interesting case-study on the origin
  of the high-ion absorption lines \sif, \cf, and \os\ in DLAs.
  The high ions in the PDLA show at least five components spanning a
  total velocity extent of $\approx$160\kms, whereas the low ions
  exist predominantly in a single component spanning just 30\kms.
  We examine various models for the origin of the high ions.
  Both photoionization and turbulent mixing layer models are
  fairly successful at reproducing the observed ionic ratios
  after correcting for the non-solar relative abundance pattern, though
  neither model can explain all five components.
  We show that the turbulent mixing layer model, in which the high ions trace
  the interfaces between the cool PDLA gas and a hotter
  phase of shock-heated plasma, can explain the average
  high-ion ratios measured in a larger sample of 12 DLAs.}
\maketitle

\section{Introduction}
In the process of galaxy assembly and evolution, interstellar and
circumgalactic gas clouds play a key role, representing
both the fuel for future star formation
and the metal-enriched by-products of past star formation.
Absorption-line spectroscopy is a powerful way to study these
gas reservoirs, to characterize their kinematic
and chemical properties, and to investigate their link to star
formation. Of particular interest for probing interstellar gas are the
damped \lya\ systems (DLAs), the highest column density absorbers defined by
log\,$N$(\hi)$>$20.3 \citep{Wo86}. Their high neutral
gas content ensures they select over-dense regions, including galaxies
and their halos. 
DLAs contain the majority of the neutral gas in the Universe in the
range $z$=0--5 \citep{Wo05}, and serve as luminosity-independent
indicators of cosmic chemical enrichment. 

Despite being known for their neutrality, DLAs
frequently show high-ion absorption overlapping in velocity space 
with the neutral lines yet tracing a separate phase (or phases) of gas.
The first high ions detected in DLAs were \cf\ and \sif\ 
\citep{Wo94, Lu95, Lu96, WP00a, WP00b}, which almost always show a
different (though overlapping) component structure than the low ions.
We have recently been surveying high-ion absorption in DLAs at
$z$=2--4 at high resolution using the VLT/UVES spectrograph, finding
the high ions are ubiquitous.
In data with S/N$\ga$30 per pixel, we report detection rates of 
$>$34\% for \os\ \citep{Fo07a}, 100\% for \cf\ and \sif\ \citep{Fo07b},
and 13\% for \nf\ \citep{Fo09}.
High-ion absorption is also seen in a composite SDSS spectrum of
several hundred DLAs \citep{Ra10}. In \citet{Fo07a} we favored the
interpretation that the \os\ absorption components in DLAs
trace warm-hot ($T\!\ga\!10^5$\,K) collisionally-ionized plasma,
rather than cool ($T\!\sim\!10^4$~K) photoionized plasma,
based on two observational findings: first, a tendency for the
individual \os\ components in DLAs to be broader 
than the \cf\ components (suggesting higher temperatures in the \os\ phase),  
and second, the failure of photoionization models to
reproduce the \os\ column densities measured in a particular DLA
studied in detail (the \zabs=2.076 DLA toward \object{Q2206--199})
without recourse to unreasonably large path lengths.
\citet{Le08} investigated the \zabs=2.377 DLA toward
\object{J1211+0422}, and similarly found that photoionization cannot
reproduce the observed \os\ column, although they also found that
single-phase collisional ionization models (equilibrium or
non-equilibrium) cannot explain the high-ion column densities in that system.

The origin of \os\ absorption in DLAs demands 
further investigation, since the observational characterization
of warm-hot gas in galaxy halos at redshifts 2--3 would provide
useful constraints on models of galaxy formation and evolution. 
Unfortunately, observations of \os\ absorption at these redshifts
are not straightforward, since the \os\ resonance doublet 
($\lambda\lambda$1031,1037) lies in the \lya\ forest and so often
suffers from severe contamination from foreground \hi\ absorbers 
\citep[e.g.][]{BT96, Fr10}. With a large enough DLA sample, one occasionally
comes across a case where the \os\ doublet lines are only partially
(rather than completely) blended, and where information on portions of
the \os\ line profiles can be recovered \citep{Fo07c, Le08, Si02}. 
Such a case (the DLA at \zabs=\zdla\ toward \qa) is discussed in this paper. 
This system has another advantage: 
although it lies at only 1\,630\kms\ from the background 
QSO (classifying it as a proximate DLA or PDLA), 
it is shielded from the QSO's ionizing radiation 
by a second absorber, a sub-DLA at \zabs=\zsub, very close in redshift
to the QSO at \zem=\zqso. 
The \hi\ column in the sub-DLA, log\,$N$(\hi)=19.00$\pm$0.15,
is high enough that the system is optically thick at the Lyman Limit,
but not at high photon energies ($E\!\ga\!75$\,eV, i.e.
$\lambda\!\la\!165$\,\AA). Therefore the 
sub-DLA acts to harden the ionizing radiation field incident on the PDLA.

PDLAs are defined as those DLAs lying within 5\,000\kms\ 
(or sometimes 3\,000\kms) of the redshift of the background QSO 
\citep{El02, El10, El11, Ru06, Rx07, Pr08}.
They are frequently excluded from studies of intervening
DLAs, due to their potentially unusual ionization conditions and
environmental effects (such as galaxy clustering near quasars).
However, the fact that the (enhanced) ionizing radiation field in
PDLAs is calculable can be turned to our advantage, and 
allows them to be used as case-studies for the origin of
high-ionization plasma in DLAs. That is the approach followed in this paper.

\section{Observations and Data Handling}
\qa\ has been observed with VLT/UVES \citep{De00} under three ESO
programs, each time with different dichroic settings and
wavelength coverage. These data were taken 
in 2000 (Dic2 412+860 setting, program P65.O-0063),
in 2002 (Dic2 390+860, 70.B-0258), and
in 2009 (Dic1 390+580, 383.A-0376),
all with a 1.0\arcsec\ slit (except the red-arm P70 observations, which
used a 0.9\arcsec\ slit).
We downloaded these data and their associated calibration files
from the ESO archive, and reduced them using the UVES pipeline
\citep{Ba00} in MIDAS, co-adding the individual exposures after
weighting by signal-to-noise. This produced a final spectrum with 6.6\kms\
resolution (FWHM) in $\approx$2\kms\ pixels covering the range 3\,286 to
10\,429\,\AA\ (with a gap between 5\,756 and 5\,837\,\AA).

Three DLAs are present along the \qa\ sight line, at 
\zabs=1.8639, 2.3745, and \zdla\ \citep{Le06} in the
vacuum heliocentric frame. In addition, a sub-DLA is detected
at \zabs=\zsub, with all the hallmarks of an intrinsic absorber (see \S6). 
The proximity (1\,630~\kms) of the 
$z$=\zdla\ DLA to the QSO redshift \zem=\zqso\ (derived in \S4) 
classifies it as a PDLA. The system has log\,$N$(\hi)=20.80$\pm$0.10 and
[Zn/H]=[\ion{Zn}{ii}/\hi]=$-$1.60$\pm$0.10 \citep{Le06}, 
a good indication of the true metallicity since Zn is typically
undepleted onto dust grains \citep[e.g.][]{Pe97}. 
In a WMAP 7-year cosmology 
\citep[$H_0$=71\,\kms\,Mpc$^{-1}$, $\Omega_{\rm M}$=0.27, 
$\Omega_{\Lambda}$=0.73;][]{Ja11},
the Hubble parameter at $z$=2.5 is $H(2.5)$=250\,\kms\,Mpc$^{-1}$
\citep[using the relation given in][]{Ph02}. 
Therefore, assuming the relative motion between the QSO and PDLA
is purely due to Hubble flow (i.e., assuming peculiar velocities are
negligible, though see Ellison et al. 2010), we
find the PDLA lies at $\approx$6.5 proper Mpc from the quasar. 

\section{\hi\ column densities in proximate DLA and sub-DLA}
In Figure 1 we show the \lya\ profiles of the two proximate
absorbers. \lya\ absorption from the sub-DLA at \zabs=\zsub\ lies in
the red wing of the \lya\ absorption from the DLA at \zabs=\zdla.
The sub-DLA is far-enough separated from the DLA that its own damping
wings are well-defined and can be fit to determine its \hi\ column density. 
These wings are visible in Figure 1 at observed wavelengths of 4250--4260 and
4265--4270\AA.
Voigt-profile fits to the two absorbers yield
log\,$N$(\hi)=20.80$\pm$0.10 in the PDLA and  
log\,$N$(\hi)=19.00$\pm$0.15 in the proximate sub-DLA. 

\begin{figure}[!ht]
\includegraphics[width=7cm,angle=270]{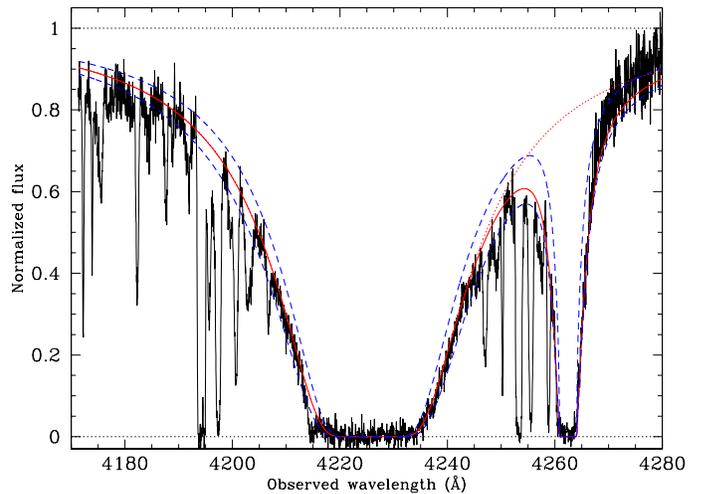}
\caption{\lya\ profile in the two proximate absorbers toward
  \qa. The red line shows the best-fit Voigt profile with
  log\,$N$(\hi)=20.80$\pm$0.10 in the DLA at \zabs=\zdla\ 
  (centre of the figure) and log\,$N$(\hi)=19.00$\pm$0.15 in the
  sub-DLA at \zabs=\zsub\ (toward the right). 1$\sigma$ errors are
  shown with blue dashed lines. The red dotted line shows the fit to
  the DLA only.}
\end{figure}

\section{Emission-line redshift and specific luminosity of Q0841+129}
No precise measurement of the redshift of \qa\ currently
exists in the literature. 
We used a flux-bisecting algorithm to measure the observed
emission-line wavelengths of  
\lyb\ 1025.722, \oi\ 1304.460 (multiplet), \sif\ 1393.755, \sif\ 1402.770, 
\nf\ 1240.14 (doublet), \cf\ 1548.195,  \hew\ 1640.42, \ct] 1908.73, and \mgw\ 2798.74.
A weighted average of the redshifts determined from each of these nine
lines gives \zem=\zqso, the value we adopt for the rest of this paper. 
Although it is common practice to apply velocity offsets to each line before
averaging \citep[e.g.][]{VB01}, we found no need to do this.
The \mgw\ redshift \citep[considered a good measure of the true
  value;][]{HW10} is 2.49512$\pm$0.00001.

We derived the specific luminosity of \qa\ at the Lyman Limit,
$L_{912}$, following \citet{Gu07}. The method uses the QSO's observed
$V$-magnitude (18.5) and emission-line redshift \zem, assumes an intrinsic QSO spectral slope
of $-$0.5 (where $F_\nu\!\propto\!\nu^{\alpha}$), and corrects for Galactic
extinction using the value $E(B-V)$=0.048 found toward \qa\ from the
maps of \citet{Sc98}. 
This gives $L_{912}$=8.86$\times10^{30}$erg\,s$^{-1}$\,Hz$^{-1}$. 
Assuming the QSO-PDLA distance is Hubble-Flow dominated (6.5\,Mpc), 
we derive a specific flux of QSO radiation at 912\,\AA\ at the PDLA of 
$F_{912}^{\rm QSO}\!\approx\!1.7\times10^{21}$erg\,cm$^{-2}$\,s$^{-1}$\,Hz$^{-1}$, 
a factor of $\approx$6 lower than the calculated flux of the UV
background (UVB) at this redshift $F_{912}^{\rm
  UVB}\!\approx\!10^{-20}$erg\,cm$^{-2}$\,s$^{-1}$\,Hz$^{-1}$
\citep{HM01, FG09}\footnote{A recent measurement of the UVB at
  $z$=2.73 using the proximity effect is $F_{912}^{\rm
  UVB}\!\approx\!4\times10^{-21}$\,erg\,cm$^{-2}$\,s$^{-1}$\,Hz$^{-1}$
  \citep{Da08},
  several times lower than the Haardt \& Madau (2001) value, but still 
  higher than the flux incident on the PDLA from the background QSO.}.  
Based on this calculation of $F_{912}$, the PDLA is outside the
proximity zone of the quasar, although for higher photon
energies, we find the PDLA does lie inside the proximity zone (see \S6).
Despite these complications in how to define proximity, we still refer
to the absorber as a PDLA, following convention. 

\section{Absorption in PDLA}
In Figure 2 we show the normalized profiles of 36 absorption lines
of both high and low ions in the PDLA.
The low-ion profiles show a simple
component structure; \sw\ $\lambda$1259 shows a single unsaturated component 
spanning $\Delta v_{90}$=30~\kms\ 
\citep[where $\Delta v_{90}$ is the width containing the central 90\% of the
  optical depth;][]{PW97}.
An additional weak low-ion component is seen in \cw\ and \siw\ near
50\kms\ in the PDLA rest-frame. Integrated column densities for the
low ions were measured using the apparent optical depth
(AOD) technique \citep{SS91}, using atomic data from \citet{Mo03},
and are presented in Table 1.

\input{tab1.tex}

\begin{figure*}[!ht]
\includegraphics[width=18cm]{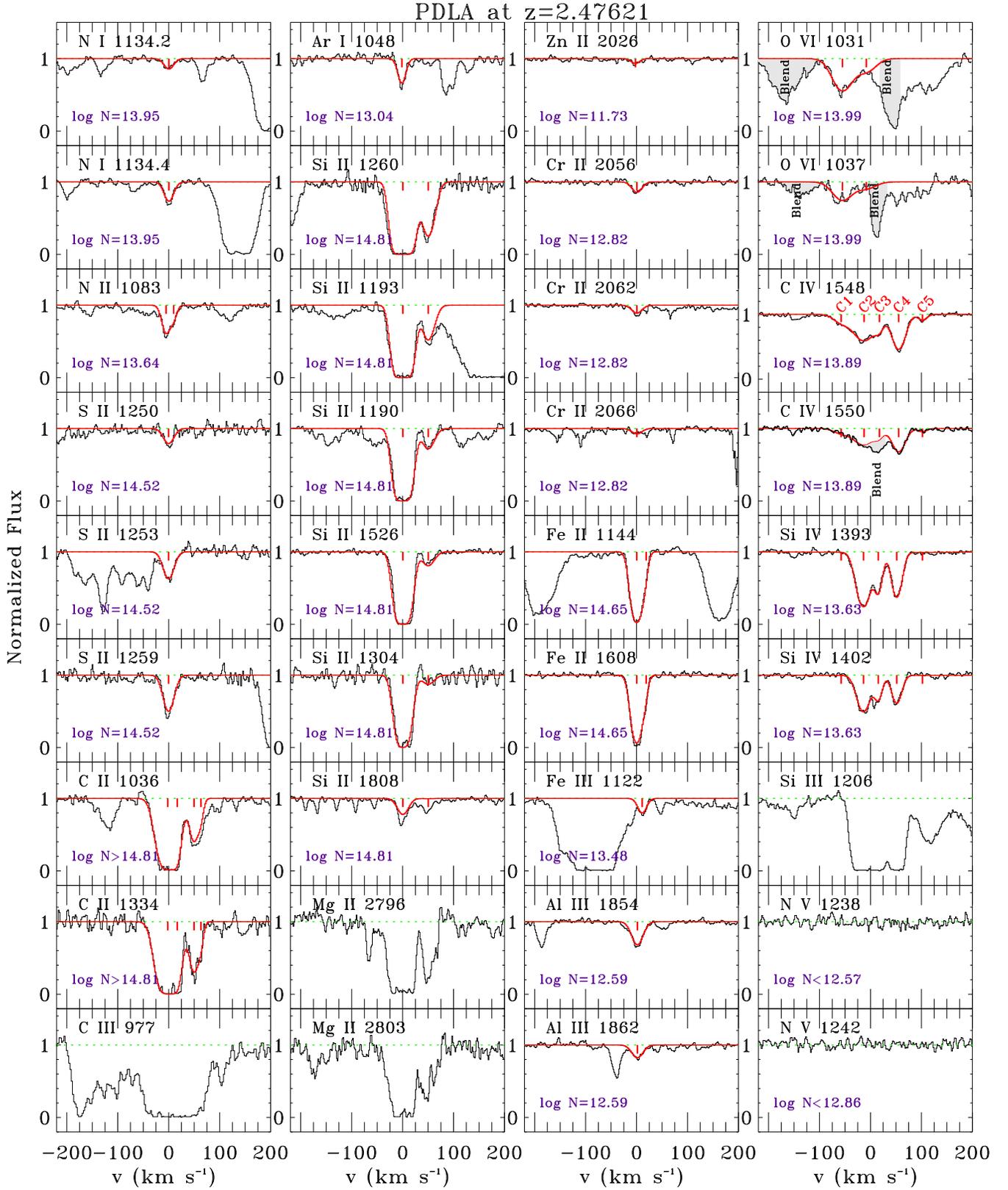}
\caption{Normalized VLT/UVES absorption-line profiles in the 
  PDLA at $z$=\zdla\ toward \qa. Red lines show Voigt-profile fits, 
  with tick marks showing component centers. Fits were not
  attempted for \sit, \ct, or \mgw\ due to extreme saturation.
  Grey shaded regions indicate blends. 
  The total column density obtained from component fitting (summed
  over components) is annotated on each panel.}
\end{figure*}

In the high ions, multi-component absorption is observed in the PDLA in
\sif, \cf, and \os, though the \os\ profiles are partly
blended. \nf\ is not detected. 
Voigt-profile fits were conducted to the high-ion absorption
using the VPFIT software package\footnote{See
  http://www.ast.cam.ac.uk/$\sim$rfc/vpfit.html}. These fits are
over-plotted on Figure 2 as red lines and detailed in Table 2, where
we also present AOD measurements of the high ions.
We chose not to tie the \cf\ and \sif\ component centroids during
the fit process, since several differences between the profiles
suggest the two ions are not fully co-spatial.

The high-ion component structure in the PDLA differs from the low-ion
structure. This is illustrated in Figure 3, where the apparent
column density profiles of several high ions and \few\
$\lambda$1608 (chosen as an unsaturated low-ion line) are compared. 
The high ions are contained in several components extending over
$\approx$160\kms, in sharp contrast to the narrowness of the \few\
line covering just $\approx$30\kms.
Based on the \cf\ profile we identify five
high-ion components in the PDLA, named as follows:
C1 centered at $-$60\kms, C2 at $-$15\kms, C3 at 15\kms, C4 at 55\kms,
and C5 at 102\kms\ (all defined relative to \zabs). 
Although \cf\ and \sif\ show similarities in their
component structure, there are clear differences:
\sif\ is detected in C2, C3, and C4, but 
not in C1 and C5. The relative strength of \sif\ to \cf\ 
changes substantially between C2, C3, and C4. Furthermore, 
C2 and C4 are broader in \cf\ than in \sif\ (see $b$-values reported
in Table 2). 

\input{tab2.tex}

\begin{figure*}[!ht]
\includegraphics[width=18.5cm]{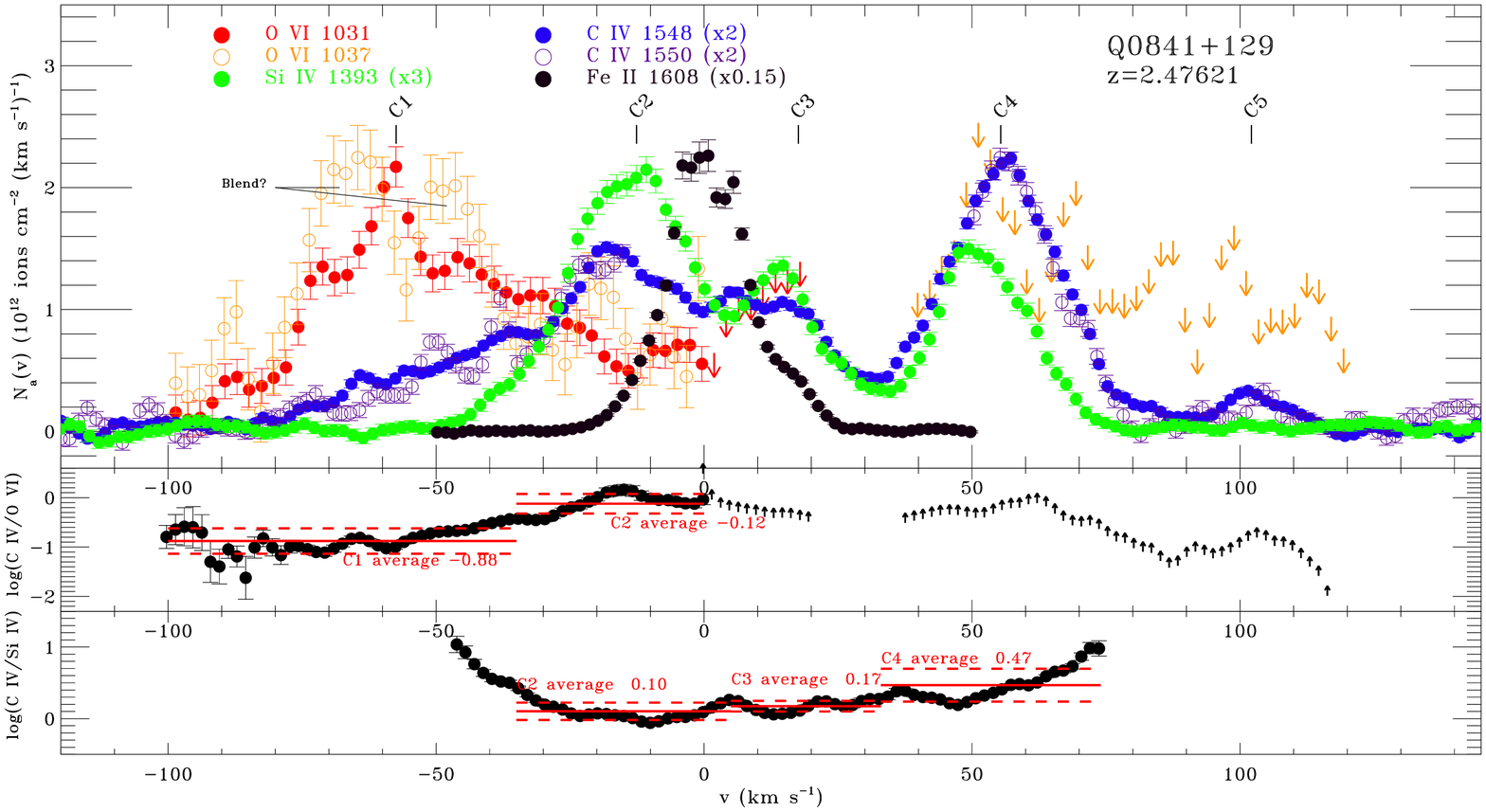}
\caption{{\bf Top panel}: Apparent column density profiles of high-ion
  and \few\ absorption in the \zabs=\zdla\ PDLA. 
  Several of the profiles have been scaled by the factors shown in the
  legend, to allow inter-comparison with the other lines.
  The \os\ profile is only shown at velocities where the two \os\
  lines overlap; upper limits for \os\ are shown at other velocities,
  using whichever of the two doublet lines gives the stronger constraint.
   {\bf Middle panel:} \cf/\os\ apparent-column-density ratio as a function of
  velocity; in pixels with upper limits on \os,
  lower limits on the ratio are given.
  {\bf Lower panel:} \cf/\sif\ apparent-column-density ratio as a function of
  velocity. The mean (solid lines) and standard deviation (dashed
  lines) of the logarithmic ratio in each component are shown in red.}    
\end{figure*}

Partial blending complicates the measurement of \os.
We report an \os\ detection in the range 
$-$100 to 0\kms\ (corresponding to components C1 and C2; see Figure 3),
because the two lines of the \os\ doublet show largely consistent optical
depth profiles at these velocities (although small
differences are observed at $-$70 to $-$40\kms). Furthermore, the integrated
AOD column densities of the two \os\ lines made over the interval
$-$100 to 0\kms\ agree to within 0.04\,dex (see Table 2).
Apparent features in \os\ $\lambda$1031
and \os\ $\lambda$1037 follow \cf\ components C3 and C4, respectively,
but in each case the data are noisy, and the other \os\ line is blended,
so we cannot confirm the detection in those components. We therefore 
only report upper limits on \os\ in C3, C4, and C5. 
For each of these three components the upper limit was derived from
whichever of the two \os\ lines gave the stronger constraint. 
These upper limits are used to constrain the high-ion ratios in each component.

The probability that the \os\ detections we report in C1 and C2 are
genuine can be roughly estimated using the known line density of the
\lya\ forest, a major contaminant. 
The data presented by \citet{Ki97} indicate a mean separation of
$\approx$500\kms\ for \lya\ forest lines with $N$(\hi)=13.1--14.3 at
$z$=2.5, indicating that the probability of a forest
line falling in a 20\kms\ interval around a given redshift
is $P_{\rm blend}\approx20/500\approx0.04$ (ignoring clustering).
The probability of two forest lines falling at that
redshift, one in each line of \os, is $P_{\rm blend}^2\approx0.0016$. 
However, the observed density of blends is far higher than one per 500\kms,
indicating that other contaminants (metal lines and higher-order Lyman
lines) are also present. We minimize the chance of these contaminants
mimicking \os\ by relying on intervals where both \os\ lines agree.

\section{Absorption in proximate sub-DLA}
Absorption-line profiles from the proximate sub-DLA at \zabs=\zsub\ 
are shown in Figure 4. This system has log\,$N$(\hi)=19.00$\pm$0.15 and
shows strong high-ion absorption and weak low-ion absorption. 
In particular, it exhibits strong multi-component \nf\
absorption, which is usually absent from intervening DLAs,
multi-component \ion{S}{vi} $\lambda$944 absorption, also unknown in
intervening DLAs, and fully saturated \os\ absorption.
It has an unusually low ratio [\oi/\hi]=$-$2.43$\pm$0.05, indicating
unusual ionization conditions given that intrinsic sub-DLA metallicities are
rarely that low \citep[e.g][]{Ku07, DZ09}.
All these properties identify the sub-DLA as an intrinsic
system arising close to the central engine of the AGN. 
The intrinsic nature of the sub-DLA supports our interpretation that
it lies between the PDLA and the QSO, as is suggested (but not
necessarily required) by the sub-DLA having a higher redshift than the PDLA.

\begin{figure}[!ht]
\includegraphics[width=8.5cm]{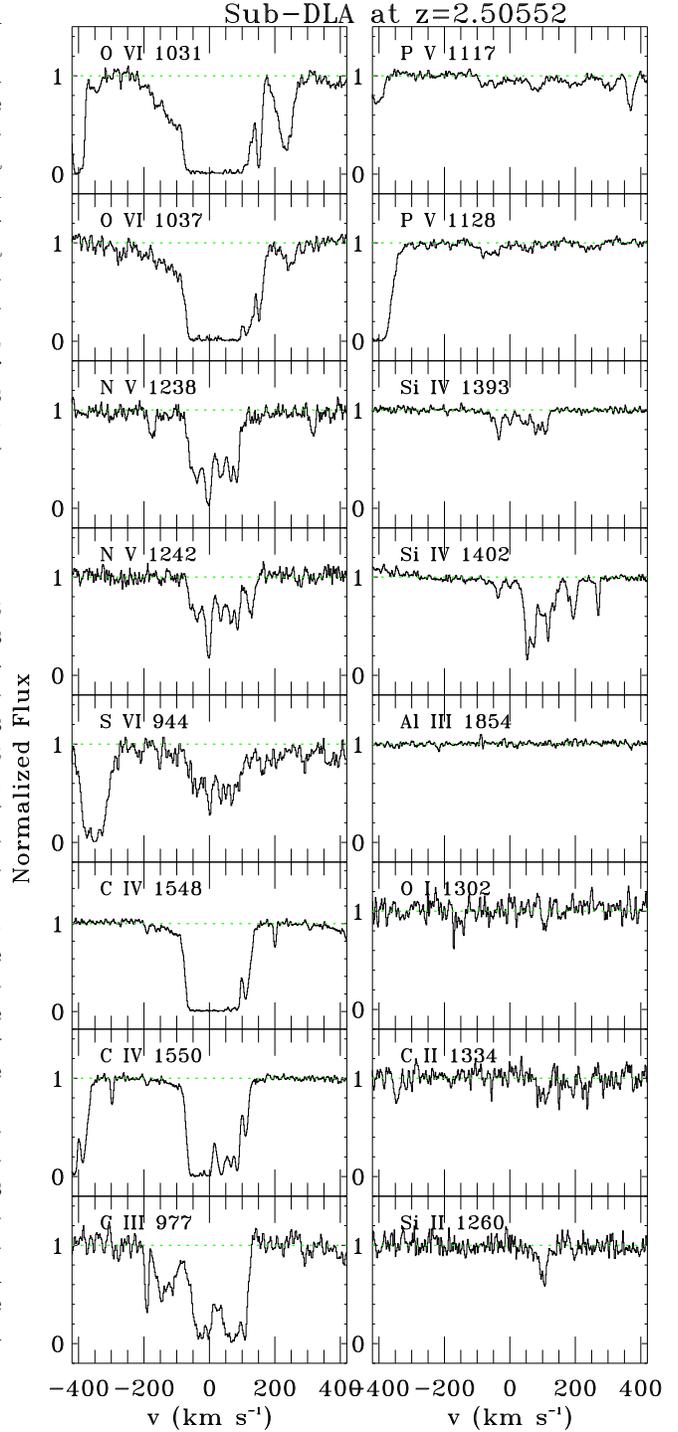}
\caption{Normalized VLT/UVES absorption-line profiles of the 
  proximate sub-DLA toward \qa. The velocity zero-point shown is
  defined by the strongest \nf\ absorption component
  ($z$=2.50552), slightly lower than the redshift defined
  by the low ions ($z$=\zsub). The presence of strong 
  \ion{S}{vi} and \nf\ absorption, and fully saturated \os\ absorption
  identifies the sub-DLA as intrinsic to the quasar.}
\end{figure}

The intrinsic sub-DLA could be used to study the physical conditions
and abundances in the gas immediately surrounding the AGN. However, in
our case, the system is of principal interest as a blocking filter, serving to
attenuate the background QSO radiation incident upon the foreground PDLA. 
The attenuating effect of the sub-DLA on the QSO's ionizing
radiation can be calculated from the following equations, which relate
the \hi\ and \hew\ optical depths to the \hi\ and \hew\ columns and
wavelengths: 

\begin{equation}
\tau^{\lambda<912}_{\rm HI}=N_{\rm HI}\,\sigma_{\rm HI}\,(\lambda/912\,{\rm \AA})^3
\end{equation}
\begin{equation}
\tau^{\lambda<304}_{\rm HeII}=N_{\rm HeII}\,\sigma_{\rm HeII}(\lambda/304\,{\rm \AA})^3
\end{equation}

where $\sigma_{\rm HI}$=6.3$\times10^{-18}$\,cm$^2$ and 
$\sigma_{\rm HeII}$=$\sigma_{\rm HI}$/4 are the
photoionization cross sections for \hi\ and \hew, respectively.
These expressions are evaluated using the 
measured sub-DLA \hi\ column log\,$N$(\hi)=19.00, assuming He/H=0.0823
by number (equivalent to a He mass fraction of 0.24), and assuming all
the He is in the form of \hew, giving
log\,$N$(\hew)=17.9. The total optical depth is then 
$\tau$=$\tau_{\rm HI}$+$\tau_{\rm HeII}$ (we do not consider
metal-line opacity).
The effect of the \hi\ and \hew\ opacity from the sub-DLA is to remove 
photons with energies $13.6\!<\!E\!\la\!75$\,eV from the
radiation field, with the \hi\ providing the opacity at
$13.6\!<\!E\!<\!54$\,eV and the \hew\ at $54\!<\!E\!\la\!75$\,eV. 
Thus the sub-DLA \emph{hardens} the ionizing radiation field incident
on the PDLA, by preferentially transmitting photons at $E\!\ga\!75$\,eV.
This is shown in Figure 5, where the total radiation field $F_\nu$
incident on the PDLA is derived. 
The radiation field is calculated as the sum of: 
(1) the quasar spectrum attenuated by the sub-DLA, assuming an intrinsic
power-law spectrum $F_\nu\propto\nu^{\alpha}$ with $\alpha$=$-$0.5,
a normalization from the value of $F_{912}$ derived in \S4, and
attenuated using the sub-DLA optical depths described above, and 
(2) the extragalactic UVB at $z$=2.5,
which is built into CLOUDY and is based on \citet{HM96, HM01}. 
We do not consider internal sources of radiation within the PDLA
(e.g. any star-forming regions), nor do we consider any attenuation of
the radiation field from the PDLA itself.
Figure 5 shows that the UVB is stronger than the QSO radiation
field at energies $E\la75$\,eV, but above this energy, the QSO
radiation dominates. The QSO radiation is therefore important to include
when investigating the origin of \os\ (which requires 114\,eV for its
creation). The combined radiation field is used as an input for the CLOUDY
modelling described in \S8.

\begin{figure}[!ht]
\includegraphics[width=8.5cm]{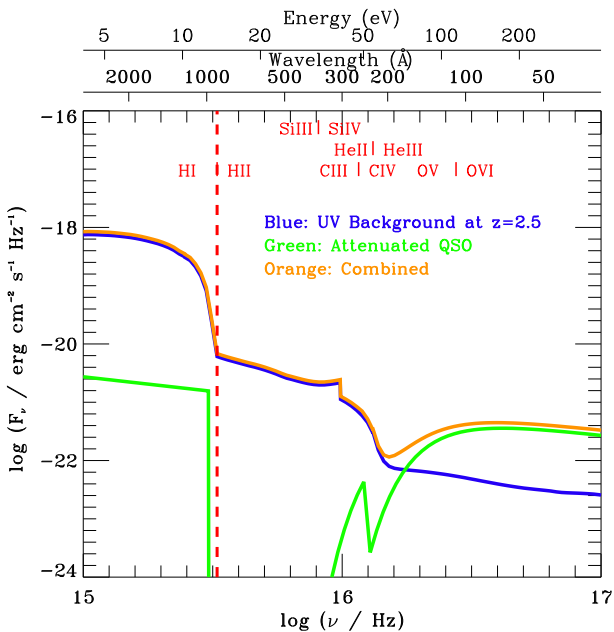}
\caption{Spectra of the ionizing radiation fields incident upon the PDLA:
  the quasar spectrum attenuated by the sub-DLA (green),
  the extragalactic UV background (UVB) at $z$=2.5 (blue), and the sum of
  the two (orange). Specific flux is plotted against frequency, 
  with the ionization edges of various ions annotated on the plot.
  The derivation of the spectra is discussed in \S6. Note how the
  QSO spectrum dominates at extreme UV energies ($E\!>\!75$\,eV).}
 \end{figure}

\section{PDLA ionization level: the low-ion phase}
The ionization level in the low-ion phase of DLAs can be diagnosed
with the \aro/\siw, \aro/\sw, \fet/\few, and \nw/\none\ ratios
\citep{Pr02, HS99, Vl01}. In the PDLA, we measure 
log\,$N$(\aro)/$N$(\siw)=$-$1.77$\pm$0.15, 
log\,$N$(\aro)/$N$(\sw)=$-$1.47$\pm$0.03,
log\,$N$(\fet)/$N$(\few)=$-$1.17$\pm$0.05
(or $-$1.02$\pm$0.05 when considering the strongest component
only), and log\,$N$(\nw)/$N$(\none)=$-$0.31$\pm$0.07, 
using the integrated VPFIT column densities. 
Each of these four ratios indicates the low-ion gas is
predominantly (but not completely) neutral with a hydrogen ionization fraction
$x$(\hi)=0.1--0.5 \citep{Pr02};
a value $x$(\hi)$>$0.5 is ruled out since
that would lead to log\,$N$(\nw)/$N$(\none)$>$--0.2 and
log\,$N$(\fet)/$N$(\few)$>$--1.0;
a value $x$(\hi)$<$0.1 is excluded since that would lead to
log\,$N$(\aro)/$N$(\siw)$>$--0.2 \citep{SJ98} and 
log\,$N$(\aro)/$N$(\sw)$>$--0.2 
(neglecting any possible dust or nucleosynthesis effects).
The \citet{Pr02} model predictions were based on CLOUDY
calculations that used the UV background as the incident 
radiation field. They are applicable to the PDLA since, for the energies of
interest for the low ions ($E\ll50$\,eV), the radiation field incident
on the PDLA is UV-background dominated (Figure 5).
Note that the measured \fet/\few\ ratio in the PDLA
is technically an upper limit as \fet\ $\lambda$1122 is detected at
low significance close to several blends, and there is an
$\approx$10~\kms\ offset between the \few\ and \fet\ velocity
centroids. However, if the feature we fit as \fet\ is a blend, 
then the true \fet\ column and \fet/\few\ ratio will be lower, thus
strengthening the conclusion that the low-ion gas is predominantly neutral.
Likewise, any contamination of \nw\ $\lambda$1083 would lower the true
\nw\ column and the \nw/\none\ ratio, again reinforcing the finding
that the low-ion gas has $x$(\hi)$<$0.5.

\section{PDLA ionization level: the high-ion phase}
In this section we use the observed high-ion column densities and
their ratios to test various models of the origin of the high-ion plasma.
The \cf/\sif, \cf/\os, and \nf/\os\ ratios measured by Voigt profile fitting in
each PDLA component are given in Table 3, with limits given where appropriate. 
We also list the average PDLA
high-ion ratios integrated over velocity, and for comparison, the
high-ion ratios measured in a range of other galaxy-halo environments. 
Considerable variation in the ratios is seen between components,
indicating variable ionization conditions;
the \cf/\sif\ ratio varies from 0.65$\pm$0.25 in C3 to $>$11.5 in C1. 
The \cf/\os\ ratio varies by over a factor of ten between the two
components where we detect \os, from 0.08$\pm$0.03 in C1 to
1.4$\pm$1.0 in C2. 

\input{tab3.tex}

\subsection{Photoionization models}
The first high-ion origin scenario we examine is photoionization (PI),
which is often considered the ionization mechanism for the \cf\
and \sif\ absorption in DLAs \citep{Lu95, MM96, Ma03, Wo05, Le08}, 
but which has been unable to explain the DLA \os\ observations presented
by \citet{Fo07a} and \citet{Le08}.
This model is of particular interest in the case of the PDLA
given the enhanced EUV radiation field incident on the absorber.
We used the CLOUDY PI code \citep[last described in][]{Fe98} to
generate models for PDLA 
components C1 and C2, the only two with \os\ measurements. 
The CLOUDY models assume the gas has uniform density
and plane-parallel geometry, and use the two-component 
(QSO+UVB) radiation field shown in Figure 5.
The model uses the measured metallicity [Zn/H]=$-$1.60
and a DLA relative abundance pattern with  
[C/O]=$-$0.5, [Si/O]=0, and [N/O]=$-$1.0, typical for DLAs with
$-2\la[$O/H$]\la-1$ \citep{Pe08, Pj08}.
Solar abundances were taken from \citet{As09}. 
No constraint on $N$(\hi) in each high-ion component is available,
since any \hi\ lines would be hidden underneath the strong \hi\ lines
from the neutral phase.
In a given component, there are two unknowns, the ionization parameter
$U\equiv n_{\gamma}/n_{\rm H}$ and the total (neutral plus ionized)
hydrogen column density $N$(H). 
We ran a grid of CLOUDY models at different values of $U$ and $N$(H), 
with log\,$U$ varied from $-$3.4 to 0.0 in 0.2\,dex intervals, and
log\,$N$(H) varied from 18.0 to 21.0 in 0.1\,dex intervals, 
and used a two-step iterative technique to solve
for the best-fit values of the two variables. 
In the first step, we found the value of $U$
that best reproduces the observed \sif/\cf\ ratio or \cf/\os\ ratio,
depending on which ions are detected.
In the second step, we found the value of $N$(H)
that reproduces the observed absolute \cf\ and \sif\ 
(or \cf\ and \os) column densities, given the value of $U$. 
The best-fit results for C1 and C2 are shown in Figure 6. 

\begin{figure}[!ht]
\includegraphics[width=8.5cm]{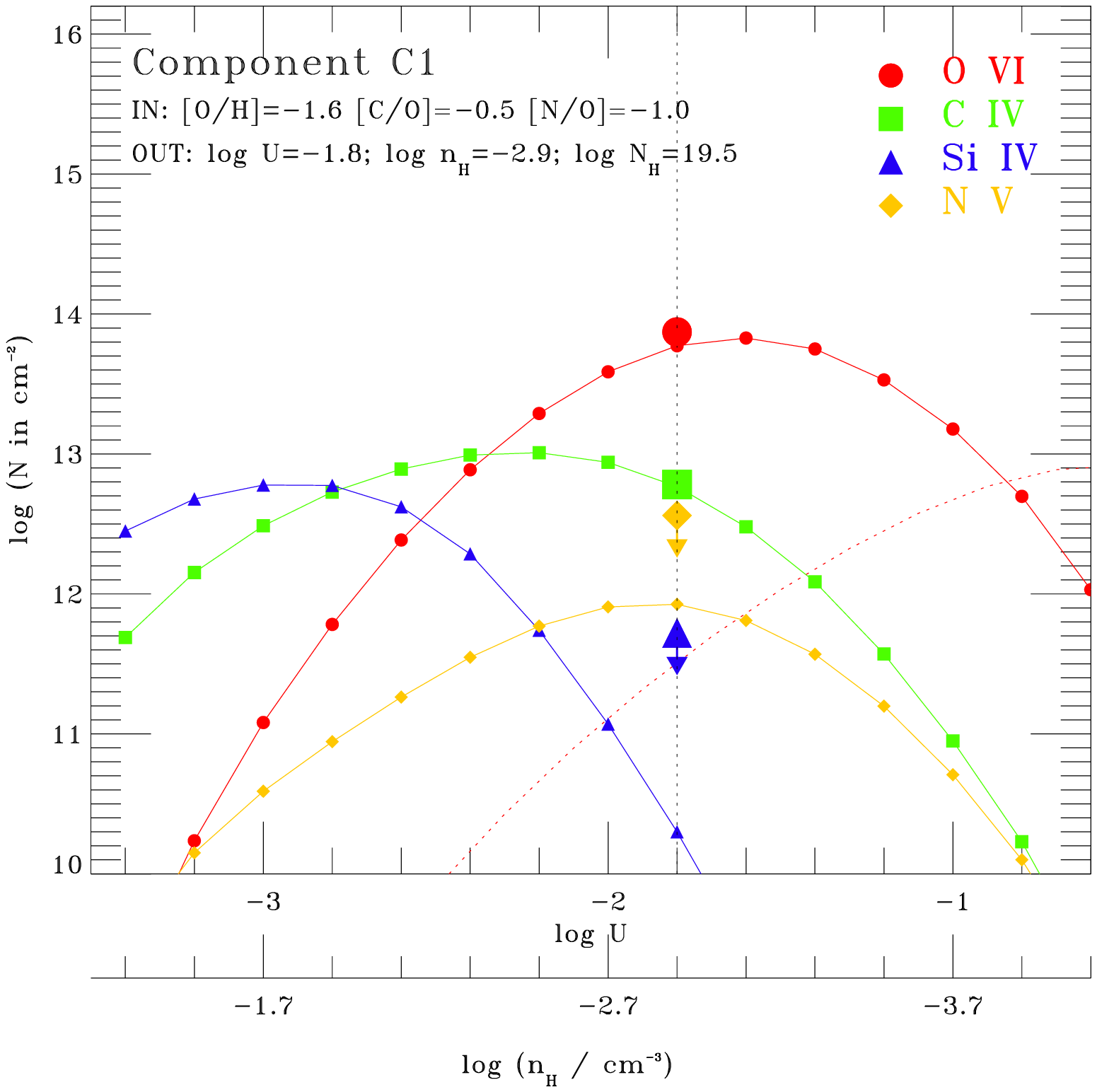}
\includegraphics[width=8.5cm]{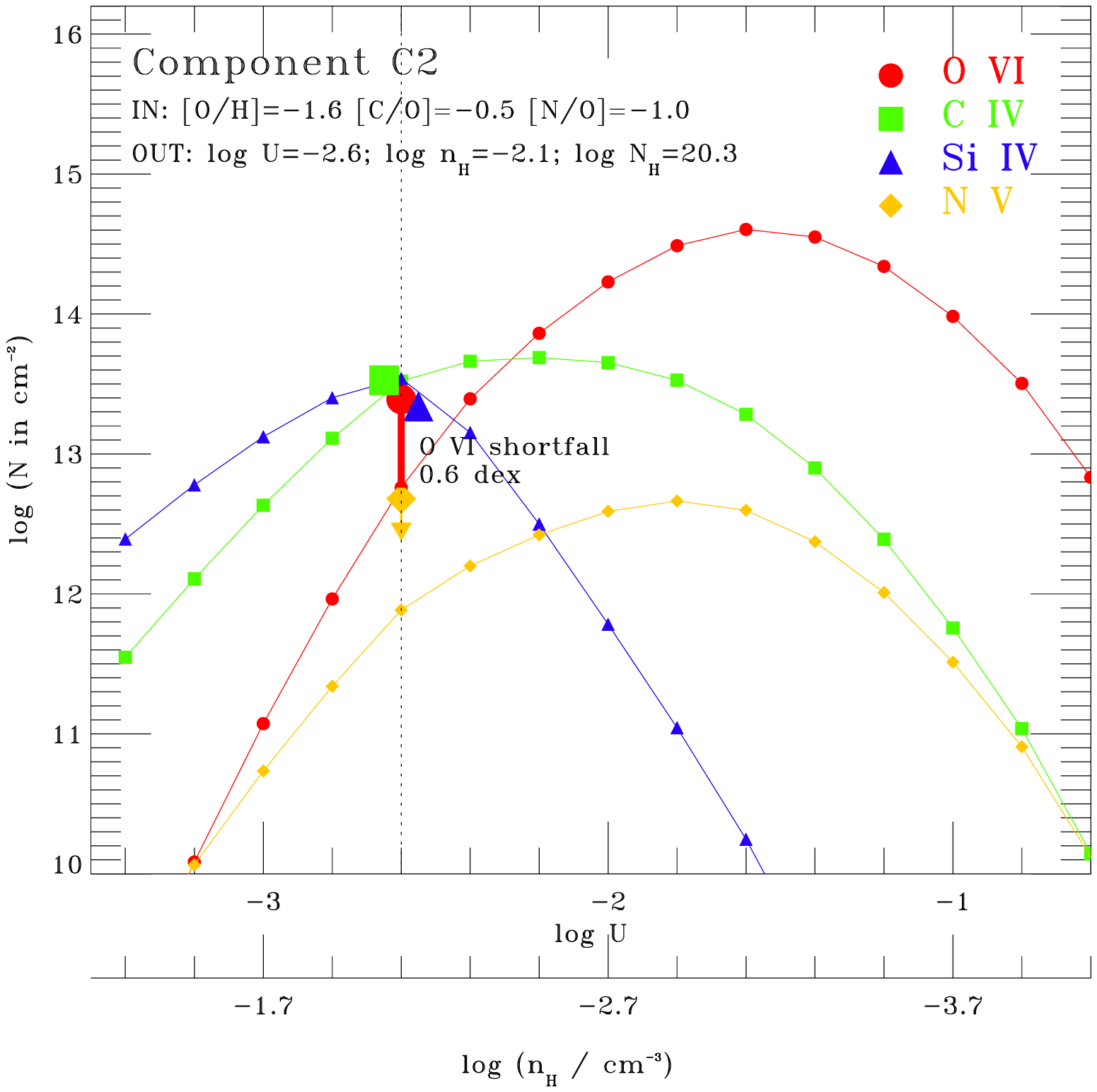}
\caption{CLOUDY PI models of the high ions 
  in the two PDLA components showing \os: 
  C1 (at $-$57\kms\ relative to \zabs), and C2 (at $-$12\kms).
  The colored lines show the predicted high-ion column densities
  versus ionization parameter log\,$U$, with the observations 
  (large data points) plotted at the best-fit value of log\,$U$. 
  The lower x-axis show the corresponding gas density $n_{\rm H}$.
  In the C1 model (upper plot), a solution is found at log\,$U$=$-$1.8
  that explains the \os, \cf, and upper limits on \sif\ and \nf.
  The dotted red line shows the \os\ prediction from a UVB-only CLOUDY
  model (i.e., without the ionizing radiation from the QSO).
  In the C2 model (lower plot), no single-phase solution is found; a
  fit to the \cf\ and \sif\ at log\,$U$=$-$2.6 under-predicts the \os\
  column by 0.6\,dex (the ``\os\ shortfall''). 
  These models are described in \S8.1.} 
\end{figure}

In C1, the strongest \os\ component where
$N$(\os)$\approx$10$N$(\cf) and where \sif\ is not detected, a PI solution  
can be found at log\,$U$=$-$1.8 and log\,$N$(H)=19.5, 
assuming [C/O]=$-$0.5. In this model the derived cloud density
log\,$n_{\rm H}$ is $\approx\!-2.9$ and the line-of-sight cloud size 
($N_{\rm H}/n_{\rm H}$) is $\approx$8\,kpc. 
This solution, found by matching the \cf/\os\ ratio, is consistent
with the non-detections of \sif\ and \nf, and
is only possible when the EUV enhancement from
the background quasar are included. UVB-only CLOUDY models do not
produce enough \os, as shown by the dotted red line on Figure 6. 

In C2, \cf, \sif, and \os\ are all detected.
In this component the \cf\ profile more closely follows \sif\ than \os. Therefore 
we use \cf\ and \sif\ as the inputs to the CLOUDY model, finding 
log\,$U$=$-$2.6$\pm$0.1, log\,$n_{\rm H}\approx-$2.1, 
and log\,$N$(H)$\approx$20.3, but this model under-predicts the observed
\os\ column by 0.6\,dex (a factor of 4). This is marked on the
lower panel of Figure 6 as the ``\os\ shortfall''. 
Therefore there is no single-phase PI solution to the
high-ion columns in this component.
Explaining the three ions in component C2 by PI alone
would require a two-phase (or multi-phase) solution, with some
fraction of the \cf\ arising with the \sif, and the remainder arising
with the \os. However, such two-phase PI models
are \emph{ad hoc}, non-unique, and without a clear physical motivation.
Indeed, \emph{any} combination of three ions (in this case \os,
\cf, and \sif) can be reproduced with a two-phase or multi-phase PI model,
so although we cannot rule out such models, we have no clear reason to
favor them.

We do not discuss PI models to C3, C4, and C5 since
\os\ measurements are unavailable in these components, and hence the
ionization state is poorly constrained. If these components are photoionized,
the implied gas densities log\,$n_{\rm H}$ derived from the \cf/\sif\
ratios are $-$1.9, $-$2.1, and $<-$2.3, respectively, assuming [C/Si]=$-$0.5. 

To summarize this sub-section, PI from the background
quasar is a viable explanation for the high ions in C1,
the PDLA component with the strongest \os,  
so long as [C/O]=$-$0.5 and we include the EUV radiation from the
background QSO. This model gives a gas density log\,$n_{\rm H}\!\approx\!-$2.9.
No single-phase PI solution is found for C2, since
the observed \os\ is four times stronger in that component than the PI
model predicts, but the \cf\ and \sif\ in C2 (and in
C3 and C4) can be explained by PI in plasma at gas densities 
log\,$n_{\rm H}\!\approx\!-2$ if [C/Si]=$-$0.5. To explore other
possibilities, we now turn to collisional ionization (CI) models.

\subsection{Collisional ionization models}
In this section several CI models are evaluated in their ability to
reproduce the high-ion observations in the PDLA: 
single-phase collisional ionization equilibrium (CIE)
and non-equilibrium models \citep{SD93, GS07},
conductive interfaces \citep[INT;][]{BH87, Bo90, Gn10}\footnote{We use
  the abbreviation INT to refer to conductive interfaces, since CI is
  already used to refer to collisional ionization.},
and turbulent mixing layers \citep[TML;][]{BF90, Sl93, Es06, KS10, Kw11}.
The common element uniting these models is that they place the 
high ions in warm-hot, collisionally-ionized plasma rather than cool
photoionized plasma. 
In the INT and TML models, the high ions trace the radiatively cooling
boundary layers between cool ($\sim$10$^4$\,K) clouds and hot
($\sim$10$^6$\,K) plasma; they differ in whether conduction or turbulent
mixing is the dominant energy transport mechanism from the hot to the
cool phase. To test the models,
we compare the high-ion column density ratios they predict with
the observations. This is shown in Figure 7, with \cf/\os\ vs \nf/\os\
plotted on the left and \cf/\os\ vs \cf/\sif\ on the right.
These ratio-ratio plots, first used by \citet{SS94},
allow observed ratios to be compared with
those measured in other environments.

Before comparing to the observations, the model predictions for the
high-ion ratios have to be corrected for the  
sub-solar C/O, C/Si, and N/O abundance ratios typical for DLAs, because
the models are computed for gas with solar relative abundances. When making
this correction, we neglect any changes in radiative cooling that result
from changing the relative elemental abundances, so we assume that the
ionic ratios scale linearly with the relative abundance of the two
elements, e.g. we assume that $N$(\cf)/$N$(\os)$\propto$(C/O),
following \citet{Fo04}.
Our adopted DLA abundances [C/O]=$-$0.5, [C/Si]=$-$0.5, and
[N/O]=$-$1.0 are typical for the 
neutral phase of DLAs \citep{Pe08, Pj08} and are assumed to apply in
the high-ion phase(s) as well. The effect of these abundance
corrections on the predicted ionic ratios is shown by the solid arrows
on Figure 7 (hereafter we refer to these new regions of ratio-ratio
space as the ``corrected predictions'' of the models).

\begin{figure*}[!ht]
\includegraphics[width=9cm]{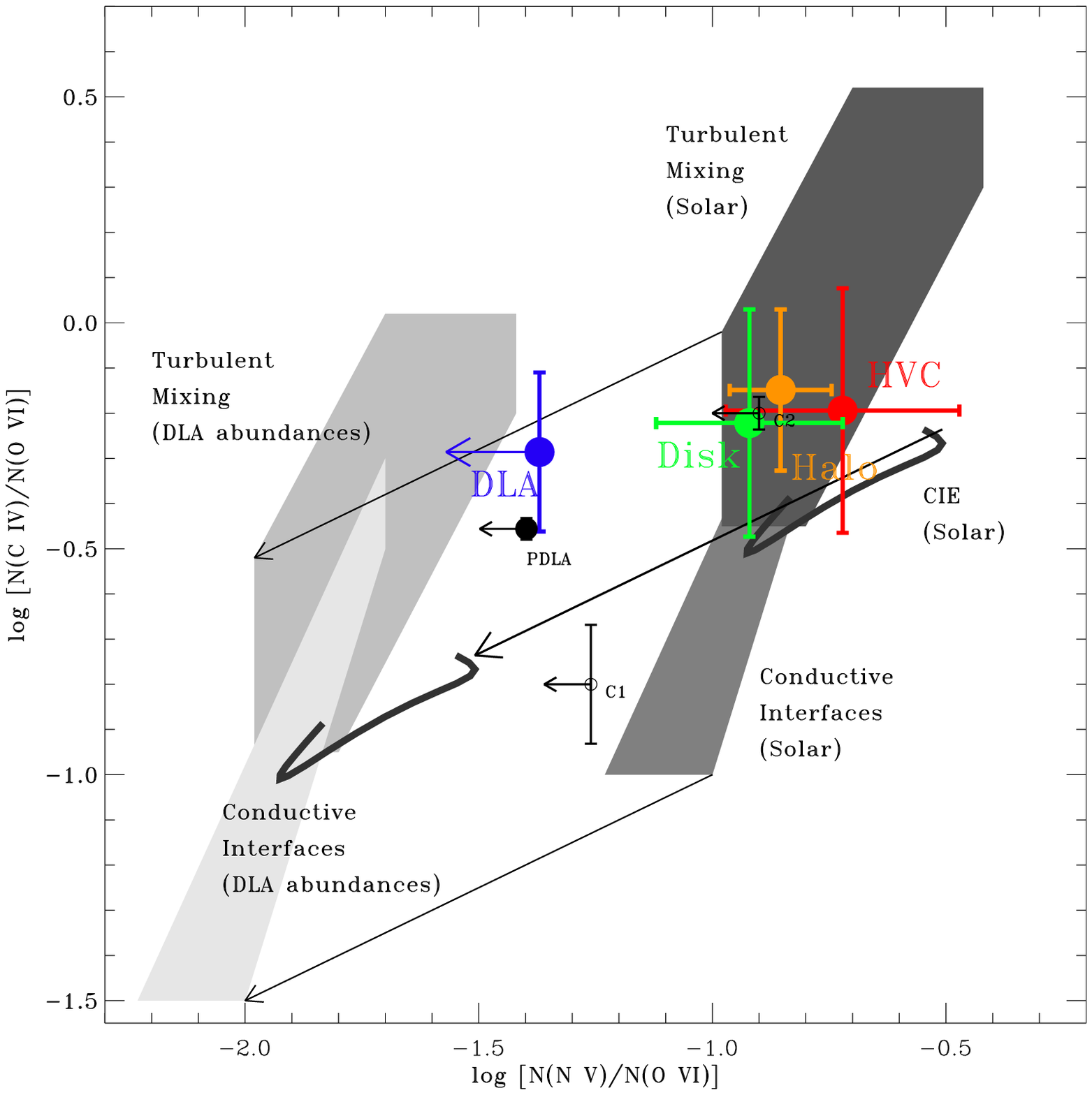}
\includegraphics[width=9cm]{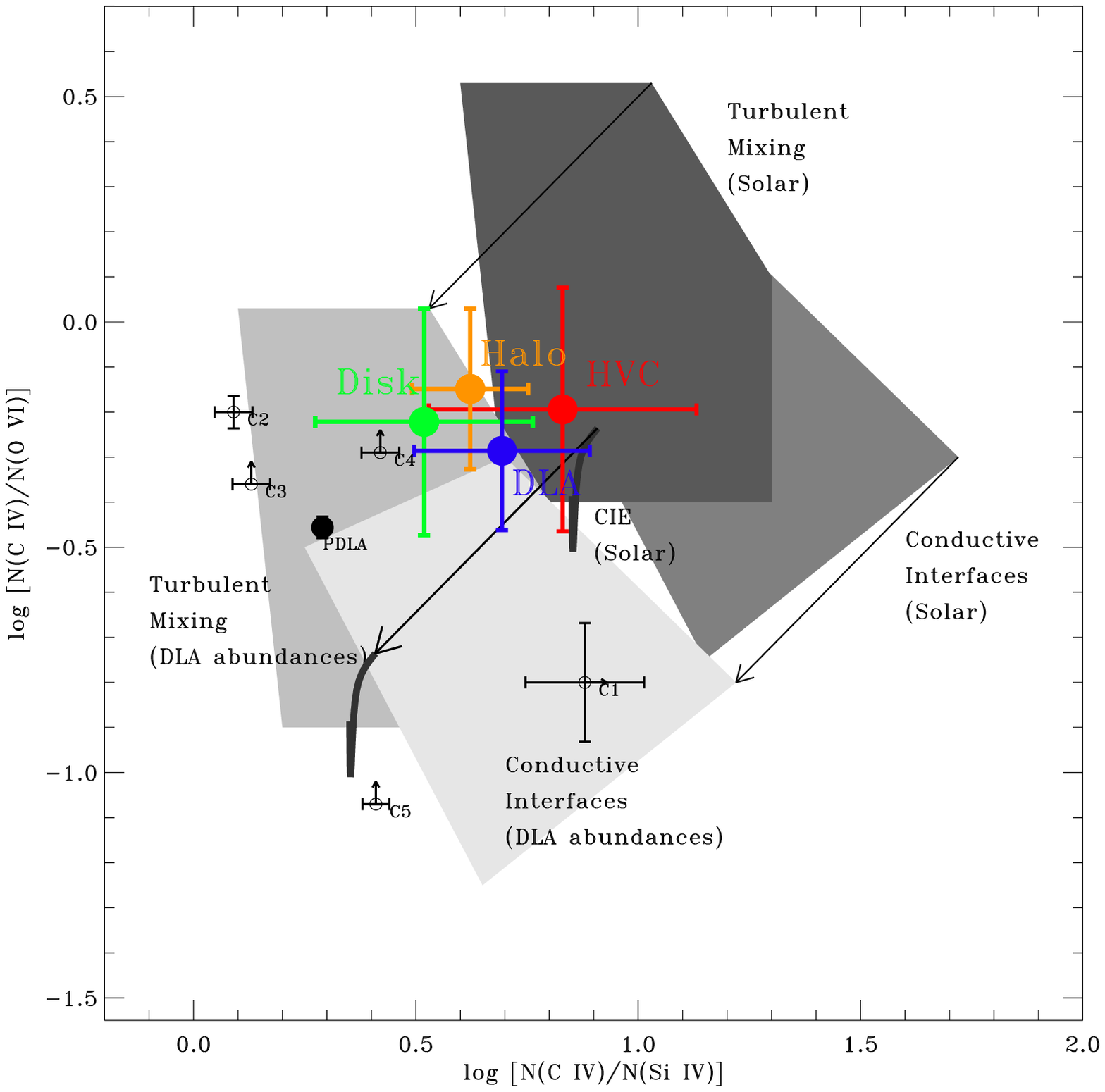}
\caption{Diagnostic ratio-ratio plots showing \cf/\os\ vs \nf/\os\
  (Fig 7a; left) and \cf/\os\ vs \cf/\sif\ (Fig. 7b; right), comparing
  observations (of this PDLA and other galaxy-halo environments)
  with ionization models. The observations shown are
  the average measurements in the \qa\ PDLA (black circles),
  the component-by-component PDLA measurements (open circles),
  the DLA average from Fox et al. (2007a) in blue, 
  the Galactic disk average from \citet{Le11} in green,
  the Galactic halo average from \citet{Wa11} in orange, 
  and the Galactic high-velocity cloud (HVC) average from \citet{Fo05} in red.
  In each case the error bar shows the standard deviation of the measurements.
  The ionization models shown (shaded regions) are
  turbulent mixing layers \citep{KS10}, conductive interfaces
  \citep{Gn10}, and CIE \citep{GS07}.
  The shaded regions show the approximate range of predicted ratios when
  allowing the model parameters to vary (see text).
  The black arrows show the effect of
  correcting the model predictions from a solar relative abundance
  pattern to a DLA relative abundance pattern with [C/O]=$-$0.5,
  [C/Si]=$-$0.5, and N/O]=$-$1.0.} 
 \end{figure*}

Our chosen TML models are those presented by \citet{KS10}.
We consider the full range of ionic ratios predicted by their
non-equilibrium Model A (the reference model, described in their
Table 1). The ratios were calculated as a function of time
in 1 Myr intervals up to a maximum mixing layer age of 80\,Myr, and
the range shown on Figure 7 reflects the variation in the ratios as
the mixing layers evolve with time.   
Our chosen INT models are those of \citet{Gn10}. We consider the
full range of column density ratios from both the ``Maximal'' and
``Central'' cases, covering all parameter choices given in their Table 4.
We show the model predictions as continuous regions of ratio-ratio
space, rather than discrete points, to account for the possibility that the
actual values of the model parameters (coronal gas temperature, density, etc)
lie between the round-number values chosen when the models were compiled.

Using the CIE models of \citet{GS07}, we find there is no
single-temperature CIE solution to the \os, \cf, and \sif\ column
densities, nor is there a single-temperature non-equilibrium CI solution in 
either the isobaric or isochoric cases (not shown on Figure 7). 
The CIE and non-equilibrium CI predictions in ratio-ratio space do
not overlap with the observed ratios -- they miss both the PDLA average and
the individual component ratios. 

On Figure 7a, inspection of the y-axis shows that the average PDLA \cf/\os\
ratio lies within the corrected range predicted by the TML models, and
toward the upper end of the corrected range predicted by the INT models. 
On the x-axis, the PDLA average \nf/\os\ ratio is an upper limit since
\nf\ is not detected, but this is consistent with the corrected
predictions of the TML and INT models. Only two individual PDLA
components (C1 and C2) are shown on this figure, since both \os\ and \nf\ are
undetected in C3, C4, and C5 so the ratio \nf/\os\ is undefined in
those components.

On Figure 7b, the average PDLA \cf/\sif\ ratio also lies
within the corrected range predicted by the INT and TML models,
although the ratios in the five individual PDLA components show a
large scatter around the average value. 
Only two of the five individual components show \cf/\os\ and \cf/\sif\
predictions that are both matched by the corrected INT prediction. 
The corrected TML model fares better, matching both ratios in four
of the five PDLA components (all apart from C1, which may be
photoionized; see \S8.1). 

The TML model of \citet{KS10} also explains the \emph{absolute} high-ion
column densities, at least for three of the components: 
the maximum \cf\ column produced per mixing layer
corrected to [C/O]=$-$0.5 is log\,$N$(\cf)=13.27, and 
we observe values in the five PDLA components of 12.40, 12.69, 12.78,
13.48, and 13.53. The final two columns are marginally 
higher than the TML model can explain.
The absolute high-ion columns are harder to explain in the
INT models of \citet{Gn10}, which predict, when corrected to [C/O]=$-$0.5,
a maximum \cf\ column per interface of log\,$N$(\cf)=12.48,
lower than the \cf\ columns observed in all PDLA components except
C1. Therefore, for the INT model to apply, unresolved multiple
interfaces would have to exist within each component, in order to
build up the observed \cf\ columns. Altogether, we conclude that the
TML model is more successful. 

It is instructive to compare the model-predicted ionic ratios to
those measured in other DLAs.
The average DLA high-ion ratios in the \citet{Fo07a} sample,
which consists of 12 DLAs with \os, are shown as large blue points on
Figure 7. These averages are successfully reproduced by both the corrected
TML and the corrected INT predictions, although only just: the TML model
cleanly reproduces the average DLA \cf/\os\ ratio ($-$0.25 in the log),
but the average DLA \cf/\sif\ ratio (0.73 in the log) is at the upper
end of the TML prediction; conversely, the INT model explains the
average \cf/\sif\ ratio but the average \cf/\os\ ratio is at the upper
end of the INT prediction. Given the uncertainties involved in
correcting the model predictions to the non-solar DLA abundance
pattern, we judge these models to be fairly successful at reproducing
the average DLA ionic ratios. 

Based on the above discussion, we conclude that boundary layers 
are a viable explanation for the \cf\ and \os\ absorption, both in
this PDLA and in DLAs in general, and that turbulent mixing is the
most successful of the models examined here, explaining the high-ion ratios
in four of the five PDLA components.  
The boundary-layer explanation is supported by the finding that 
the average \cf/\os\ and \cf/\sif\ ratios observed in DLAs are similar to those
measured in the disk, halo, and high-velocity clouds (HVCs) of the
Milky Way (a DLA itself), where interfaces and turbulent mixing are
widely invoked to explain the high ions 
\citep[e.g.][]{Sa94, Sa03, Sp96, Se03, Co04, Co05, Fo04, Fo05, Le11, Wa11}, 
although inspection of Figure 7b shows that the Galactic disk and halo average 
\cf/\sif\ ratios are marginally
lower than the predictions of the solar-abundance TML and INT models shown.
The DLA average \nf/\os\ ratio is at least 0.5\,dex lower than the Galactic 
disk and halo average \nf/\os\ ratios, but this can be explained by the
sub-solar N/O ratio present in DLAs, i.e. as an abundance effect \citep{Fo09}.

The boundary-layer scenario, in which the high ions
are cooling rather than photoionized,
has other advantages: it explains the metallicity-independence of the
strength of \os\ absorption in galaxy halos \citep{He02, Fo11}, and
fits naturally into the established picture of galaxy formation 
where cold streams pass through hot plasma,
since boundary layers will naturally arise between the two phases.
Indeed, in this picture, the \os\ and \cf\ detections in DLAs
indicate (without tracing directly) the presence of a
hot ($\sim$10$^6$\,K) phase of surrounding plasma in the host galaxies.
Such hot galactic coronae are predicted
to arise by shock heating of gas falling into potential wells
\citep[hot-mode accretion;][]{RO77, BD03, Ke05, Ke09}, and by feedback
from stellar activity (supernovae) within the galaxies themselves
\citep[e.g.][]{Ra09, Cr10, He10}. 
If this boundary layer scenario is correct, 
then the larger significance of \os\ observations in DLAs at $z$=2--3
would then be that feedback and/or hot-mode accretion
are already active at this epoch and have created hot galactic coronae. 
Further studies of intervening DLAs where the high-ion
ratios can be measured component-by-component,
and where the ionizing radiation field and the influence of photoionization
are expected to be lower, will confirm whether this picture is correct. 

Finally, we note that neither the boundary-layer scenario nor the
photoionization scenario explains the high-ion kinematics, which are
considerably more extended than the low-ion kinematics, both in the
PDLA and in DLAs in general \citep{WP00a, Fo07b}. This indicates that
an energetic mechanism (e.g. a galactic wind or galaxy-galaxy
encounter) has dispersed the high ions over a large velocity interval.
Simulations of the high-ion velocity fields in DLAs are needed to
investigate the nature of this mechanism.

\section{Summary}
In order to explore the processes that generate high-ion plasma in
high-redshift galaxy halos, we have presented new VLT/UVES
high-resolution observations of the 
PDLA at \zabs=\zdla\ toward \qa, focusing on the origin of the
high-ion absorption. This system lies at 1\,630\kms\ from the 
background QSO, has log\,$N$(\hi)=20.80$\pm$0.10,
a metallicity [Zn/H]=$-$1.60$\pm$0.10,
a low-ion velocity width $\Delta v_{90}$=30~\kms, and
a hydrogen ionization fraction in the low-ion phase of 0.1--0.5 as
determined from the \aro/\siw, \aro/\sw, \nw/\none\ and \fet/\few\ line ratios.
The high ions in the PDLA show a multi-component structure;
at least five \cf\ components are seen spanning a total velocity extent of
$\approx$160\kms\ with $b$-values ranging from 8.8$\pm$1.8\kms\ to
27.3$\pm$3.1\kms. \sif\ is detected in three of these components; \os\
is detected in two, and blended in the others.

The PDLA is (partly) shielded from the QSO's ionizing radiation by an
intrinsic sub-DLA at \zabs=\zsub, very close to the QSO emission-line
redshift of \zem=\zqso. This filtering arrangement hardens the
ionizing radiation field incident on the PDLA, allowing us to test
whether photoionization (PI) is a viable origin mechanism.
We find that a single-phase PI model at log\,$n_{\rm H}\!\approx\!-2.9$
can successfully explain one of the two components showing \os\ (C1),
so long as [C/O]=$-$0.5, but no single-phase PI
model is found for the other component showing \os\ (C2),
since the \os\ in this component is a factor of four stronger than
the model predicts. The \cf\ and \sif\ profiles in C2, C3, and C4, are
consistent with PI at log\,$n_{\rm H}\!\approx\!-2$ if [C/Si]=$-$0.5.
Single-phase collisional ionization models (equilibrium or
non-equilibrium) are unable to reproduce the high-ion observations.
 
An alternative model, where the high ions trace
the turbulent mixing layers between the warm DLA gas and a
hot exterior medium \citep{KS10}, can successfully reproduce the ionic ratios in 
four of the five components (C2, C3, C4, and C5). This model can also explain
the average high-ion ratios \cf/\os, \nf/\os, and \cf/\sif\ observed
in a larger sample of 12 DLAs. 
Turbulent mixing layers are often invoked (along with conductive
interfaces) to explain high-ion observations in the halo of the Milky Way (itself a DLA).
If correct, this model provides indirect evidence for the existence of a
hot phase of coronal plasma in and around DLA galaxies. 
Further observations of high-ion absorption in individual intervening
DLAs are needed to test this model.

\begin{acknowledgements}
AJF acknowledges support from an ESO Fellowship,
useful conversations with Bob Carswell, Paul Hewett, and Sara Ellison,
and constructive comments from the anonymous referee. 
He is grateful to Kyujin Kwak for providing his Myr-by-Myr 
TML predictions, and to Bart Wakker for advice on CLOUDY models.
PPJ and RS gratefully 
acknowledge support from the Indo-French Center for Promotion of
Advanced Research (Center Franco-Indien pour la Promotion de la
Recherche Avanc\'ee) under contract No. 4303.
\end{acknowledgements}

\end{document}

%% file: tab1.tex
\begin{table}
\begin{minipage}[t]{8.5cm}
\caption{Apparent optical depth measurements of low-ion absorption in the Q0841+129 PDLA}
\centering \begin{tabular}{lccc} \hline \hline
Line & $v_-$,$v_+$\footnote{Velocity range used in AOD integration, relative to \zabs=2.47621. The default is $-$30 to 30\kms, but this is reduced when necessary to avoid blends.} & $W_\lambda^{\rm r}$\footnote{Rest-frame equivalent width, with error that includes statistical and continuum-placement contributions combined in quadrature with a 2.5\,m\AA\ error accounting for the uncertainties in the velocity integration range.} & log\,$N_{\rm a}$\footnote{Apparent column density. Lower limit given if line is saturated. Errors include statistical and continuum-placement contributions combined in quadrature with a 0.01\,dex error accounting for the uncertainties in the velocity integration range. Entries marked $^*$ are potentially blended and should be treated with caution.}\\
    & (\kms) & (m\AA) & ($N_{\rm a}$ in \sqcm) \\
\hline
  \none\ 1134.2 &    $-$30, 30 &           13$\pm$3 &     13.96$\pm$0.03     \\
  \none\ 1134.4 &    $-$30, 30 &           26$\pm$3 &     13.99$\pm$0.01     \\
      \nw\ 1083 &    $-$30, 30 &           44$\pm$3 &     13.69$\pm$0.01$^*$ \\
      \sw\ 1250 &    $-$30, 30 &           22$\pm$3 &     14.51$\pm$0.02     \\
      \sw\ 1253 &    $-$30, 30 &           49$\pm$3 &     14.57$\pm$0.01     \\
      \sw\ 1259 &    $-$30, 30 &           59$\pm$3 &     14.53$\pm$0.01     \\
      \cw\ 1036 &    $-$30, 30 &          180$\pm$3 &           $>$14.72     \\
      \cw\ 1334 &    $-$30, 30 &          243$\pm$3 &           $>$14.65     \\
       \ct\ 977 &    $-$30, 30 &          200$\pm$3 &           $>$14.17     \\
     \aro\ 1048 &    $-$30, 30 &           24$\pm$3 &     13.05$\pm$0.02$^*$ \\
     \siw\ 1260 &    $-$30, 30 &          241$\pm$3 &           $>$13.85     \\
     \siw\ 1193 &    $-$30, 30 &          207$\pm$3 &           $>$14.06     \\
     \siw\ 1190 &    $-$30, 30 &          179$\pm$4 &           $>$14.27     \\
     \siw\ 1526 &    $-$30, 30 &          220$\pm$3 &           $>$14.52     \\
     \siw\ 1304 &    $-$30, 30 &          173$\pm$3 &           $>$14.59     \\
     \siw\ 1808 &    $-$30, 30 &           54$\pm$3 &     14.99$\pm$0.01$^*$ \\
     \mgw\ 2796 &    $-$30, 30 &          505$\pm$3 &           $>$13.60     \\
     \mgw\ 2803 &    $-$30, 30 &          494$\pm$4 &           $>$13.92     \\
     \znw\ 2026 &    $-$30, 30 &           12$\pm$3 &     11.83$\pm$0.04     \\
     \crw\ 2056 &    $-$15, 15 &           21$\pm$3 &     12.76$\pm$0.02     \\
     \crw\ 2062 &    $-$15, 15 &           20$\pm$3 &     12.87$\pm$0.02     \\
     \crw\ 2066 &    $-$15, 15 &           10$\pm$3 &     12.74$\pm$0.03     \\
     \few\ 1144 &    $-$30, 30 &          119$\pm$3 &           $>$14.34     \\
     \few\ 1608 &    $-$30, 30 &          161$\pm$3 &           $>$14.51     \\
     \fet\ 1122 &    $-$ 5, 22 &           21$\pm$3 &     13.58$\pm$0.01$^*$ \\
     \alt\ 1854 &    $-$20, 20 &           56$\pm$3 &     12.59$\pm$0.01     \\
     \alt\ 1862 &    $-$20, 20 &           30$\pm$3 &     12.60$\pm$0.01     \\
     \sit\ 1206 &    $-$30, 30 &          249$\pm$3 &           $>$13.75     \\
\hline
\end{tabular}
\end{minipage}
\end{table}

%% file: tab2.tex
\begin{table*}
\begin{minipage}[t]{18cm}
\caption{Voigt-profile fits and AOD measurements of high-ion absorption in the PDLA toward Q0841+129}
\centering \begin{tabular}{lcccc cccc} \hline \hline
Ion & Comp. & $v_0$\footnote{Velocity centroid relative to \zabs=2.47621.} & $b$ & log\,$N$ & log\,$N_{\rm tot}$\footnote{Total VPFIT column density, summed over all components, with errors added in quadrature.} & log\,$N_{\rm a}$(strong)\footnote{Apparent column density measured from strong and weak doublet lines in velocity interval indicated.} & log\,$N_{\rm a}$(weak)$^{\rm c}$ & $v_-$,$v_+$\footnote{Velocity range used in AOD integration, relative to \zabs.}\\
    & &  (\kms) & (\kms) & ($N$ in \sqcm) & ($N$ in \sqcm) & ($N$ in \sqcm) & ($N$ in \sqcm) &  (\kms) \\
\hline
  \sif & C1 &                  ... &                  ... &             $<$11.72 &       13.63$\pm$0.01 &       13.60$\pm$0.01 &                           13.63$\pm$0.01 &             $-$50,80 \\
   ... & C2 &          $-$13$\pm$1 &         16.7$\pm$0.4 &       13.34$\pm$0.01 &                  ... &       13.28$\pm$0.01 &                           13.28$\pm$0.01 &             $-$100, 0 \\
   ... & C3 &             15$\pm$1 &          9.9$\pm$0.6 &       12.88$\pm$0.02 &                  ... &                  ... &                                      ... &                      \\
   ... & C4 &             51$\pm$1 &         12.5$\pm$0.3 &       13.06$\pm$0.01 &                  ... &                  ... &                                      ... &                      \\
   ... & C5 &                  ... &                  ... &             $<$11.86 &                  ... &                  ... &                                      ... &                      \\
   \cf & C1 &          $-$57$\pm$3 &         19.2$\pm$3.2 &       12.78$\pm$0.13 &       13.89$\pm$0.03 &       13.89$\pm$0.01 &                                (blended) &           $-$100,120 \\
   ... & C2 &          $-$12$\pm$1 &         27.3$\pm$3.1 &       13.53$\pm$0.04 &                  ... &       13.49$\pm$0.01 &                           13.49$\pm$0.01 &             $-$100, 0 \\
   ... & C3 &             17$\pm$1 &          8.8$\pm$1.8 &       12.69$\pm$0.14 &                  ... &                  ... &                                      ... &                      \\
   ... & C4 &             55$\pm$1 &         15.8$\pm$0.3 &       13.48$\pm$0.01 &                  ... &                  ... &                                      ... &                      \\
   ... & C5 &            102$\pm$1 &          9.4$\pm$1.5 &       12.40$\pm$0.05 &                  ... &                  ... &                                      ... &                      \\
                                                                                                                                                    \os\footnote{Voigt fit for \os\ only possible in components C1 and C2. \os\ AOD measurements made over velocity region where the profiles of the two doublet lines overlap. The \cf, \sif, and \nf\ measurements are repeated over this interval so the ionic ratios can be derived.} & C1 &          $-$55$\pm$3 &         25.0$\pm$2.5 &       13.87$\pm$0.06 &       13.99$\pm$0.11 &       13.97$\pm$0.01 &                           13.93$\pm$0.01 &            $-$100, 0 \\
                                                                                                                                                                                                                                                                                                                                                                                                                                      ... & C2 &          $-$ 8$\pm$5 &        24.3$\pm$17.5 &       13.39$\pm$0.32 &                  ... &                  ... &                                      ... &                      \\
\nf\footnote{No detection in either \nf\ line. Upper limits are 3$\sigma$.} & C1 & ... & ... & ... & ... & $<$12.57 & $<$12.86 & $-$100, 0\\
\hline
\end{tabular}
\end{minipage}
\end{table*}

%% file: tab3.tex
\begin{table}
\begin{minipage}[t]{8.5cm}
\caption{Component high-ion ratios: PDLA vs averages in other galaxy-halo environments}
\centering \begin{tabular}{lccc} \hline \hline
Component (velocity) & $\frac{N(\cf)}{N(\sif)}$\footnote{Component ratios derived from VPFIT columns, with limits given for components with \sif\ and \nf\ non-detections, or potential \os\ blending.} & $\frac{N(\cf)}{N(\os)}$ & $\frac{N(\nf)}{N(\os)}$\\
\hline
C1  (                         $-$57\kms) &        $>$ 11.5 &                                      0.08$\pm$0.03 &   $<$0.049 \\
C2  (                         $-$12\kms) &   1.54$\pm$0.17 &                                        1.4$\pm$1.0 &   $<$0.194 \\
C3  (                            17\kms) &   0.65$\pm$0.25 &                                            $>$0.25 &        ... \\
C4  (                            55\kms) &   2.64$\pm$0.07 &                                            $>$0.56 &        ... \\
C5  (                           102\kms) &        $>$  3.4 &                                            $>$0.06 &        ... \\
PDLA average\footnote{PDLA average ratios derived from AOD integrations. The \cf/\sif\ ratio is measured over the full interval $-$100 to 120\kms. The \cf/\os\ and \nf/\os\ ratios are measured over the (unblended) velocity interval $-$100 to 0\kms.} & 1.95$\pm$0.09 & 0.35$\pm$0.02 & $<$0.042\\
\hline
                                  DLA average\footnote{Average and standard deviation of ratio among 12 DLAs with \os\ absorption at $z$=2--3 \citep{Fo07a}. The \nf/\os\ average is an upper limit since \nf\ is only detected in 3 of these 12 DLAs.} & 5.37$\pm$2.20 & 0.56$\pm$0.28      & $<$0.046 \\
Milky Way disk average\footnote{Taken from \citet{Le11}. Based on measurements made along 38 Galactic disk sight lines.} & 
3.3$\pm$2.5 & 0.60$\pm$0.47 & 0.12$\pm$0.07\\
Milky Way halo average\footnote{Taken from \citet{Wa11}. Based on measurements made along 58 extragalactic sight lines.} & 
4.19$\pm$1.47 & 0.71$\pm$0.36 & 0.14$\pm$0.04\\
LMC average\footnote{Range observed in four LMC sightlines \citep[not including \hw\ regions;][]{LH07}.} & 
1.5--2.5 & $<$0.09--0.59 & $<$0.33\\
\hline
\end{tabular}
\end{minipage}
\end{table}